\documentstyle[aps,epsf,floats]{revtex}
\begin{document}
\draft
\title{Scalar-isoscalar mesons in pion production at threshold}
\author{C. M. Maekawa}
\address{Instituto de F\'{\i}sica Te\'orica, 
Universidade Estadual Paulista\\ 
Rua Pamplona 145, 01405-900, S\~ao Paulo, SP, Brazil}
\author{C. A. da Rocha}
\address{Departamento de Desenvolvimento Tecnol\'ogico, 
Universidade S\~ao Judas Tadeu\\
Rua Taquari 546, 03166-000, S\~ao Paulo, SP, Brazil}
\date{July 4, 1999}
\maketitle

\begin{abstract}
We present the contributions of the non-linear chiral scalar field $S$ 
to cross sections of  $pp\rightarrow d\pi ^{+}$, 
$pp\rightarrow pn\pi ^{+}$ and  $pp\rightarrow pp\pi ^0$ 
reactions at threshold. 
We compare our results with the meson $\sigma$ contribution from the 
$\sigma$-linear model. 
We show that the chiral scalar field $S$ is almost 5 times bigger 
than the other contributions, which 
indicates that the two-pion exchange dynamics embebbed in the $S$ field
is the most important term for the pion production at threshold.
\end{abstract}

The 90's decade presented the advent of new accelerators 
and very precise detecting systems, which led to good and 
accurate data for $np\rightarrow d \pi^0$ \cite{Hut+91},
$p p\rightarrow p p \pi^0$ \cite{Mey+90,Bon+95}, 
$p p \rightarrow d \pi^+$ \cite{Dro+96,Hei+96},
and $p p \rightarrow pn \pi^+$ \cite{Dae+95,Dae+98} reactions
near threshold. 
Before these new data, nuclear S-wave pion production was usually 
described by a single nucleon (impulse approximation term) and a pion 
rescattering mechanism (seagull term) \cite{KR66}.
However, these new data proved that the impulse approximation and the pion
rescattering term were not enough to take into account the magnitude of 
the new cross sections \cite{MS91}, underestimating them by a factor of 5. 
To fix this problem, many authors suggested the use of 
short range dynamics \cite{LR93,HMG94,Han+95} through the exchange of heavy
mesons like $\sigma$ and $\omega$, which proved that a scalar-isoscalar
meson has a critical role in $\pi^0$ production. 
However, this procedure presented
a lot of theoretical uncertainty embebbed in the coupling constants, 
which motivates Cohen {\it et al.} \cite{Coh+96} to use Chiral Perturbation
Theory ($\chi_{PT}$) to study the $\pi^0$ production. The main idea was to 
organize the several potentially significant mechanisms of pion production.
They adapted the power counting and estimated leading and next-to-leading 
contributions, but the final result was again a factor of 5 below data.
Other $\chi_{PT}$ calculations were performed, stressing the importance of
(a) the rescattering term \cite{SC95} and (b) the effect of loops
\cite{SC95,Han+98,moalem,Dmi+99}. The main problem in these evaluations is
the large number of contributions, especially in the loop case, which 
turns the result to have some theoretical uncertainty.
Therefore it would be very useful if we could take into account the 
intermediate range contribution of the two-pion loop in some effective 
and easier manner. The most well known procedure to do that is to 
represent the pionic loop by an effective scalar-isoscalar meson, as done 
by several authors regarding the $NN$ interaction \cite{TRS75,Mac89} and, 
in this paper, we are looking close at contribution of such scalar-isoscalar
meson exchange between nucleons to the cross section of pion production at 
threshold. Such scalar-isoscalar meson has been used to improve the contribution of 
impulse and rescattering terms \cite{LR93,HMG94,Han+95}. 

There are two models for scalar-isoscalar meson: $\sigma $-linear model and
a scalar field $S$ implemented in the framework of nonlinear Lagrangian
which is chiral invariant and appears naturally when nonlinear fields are
obtained from linear ones \cite{Wei 67}. Comparison between chiral $S$ and 
$\sigma $-linear in three-body forces \cite{Mae+98} shows a favorable 
result to the former. 

In the framework of $\sigma -$linear model the dynamics 
comes from the following Lagrangian
\begin{equation}
L_\sigma =-g\bar N\left( \sigma +i\vec \tau \cdot \vec \pi \gamma _5\right)
N\,, 
\label{Eq.1}
\end{equation}
where $\vec \pi $ is the pion field, $N$ is the nucleon field that 
transforms linearly.
The non-linear Lagrangian for the scalar-isoscalar meson reads
\begin{equation}
L_S^{PV}=\frac g{2m}\bar \psi \gamma _\mu \gamma _5\vec \tau \psi \cdot
D^\mu \vec \phi -g_SS\bar \psi \psi \,,
\label{Eq.2}
\end{equation}
and equivalent results to (\ref{Eq.2}) can be obtained from the following
pseudoscalar form 
\begin{equation}
L_S^{PS}=-\left( g+\frac{g_S}{f_\pi }S\right) \bar N\left( f\left( \vec \phi
\right) +i\vec \tau \cdot \vec \phi \gamma _5\right) N+\cdots.  \,,
\label{Eq.3}
\end{equation}
where  $f_\pi $, $g$ and $g_S\,$ are, respectively, pion decay constant,
$\pi N$ coupling constant$\,$ and $SN\,$ coupling constant, $\vec \phi $ is
the pion field in nonlinear realization and  
$f\left( \vec \phi \right) =\sqrt{f_\pi ^{\;2}-\vec \phi ^2}$ \ ; in ``$\ldots$''
we subsume all other possible interactions that contribute to the pion 
production, but are not required here since they are of lower magnitude. 
The pion-nucleon decay constant $f_\pi$ 
relates to $g$ by the Goldberger-Treimann (GT) relation, $g\,f_\pi=g_A\, m$, 
where $g_A$ is the axial
pion-nucleon coupling constant and $m$ the nucleon mass.
This is a phenomenological requirement since the Lagrangians in 
Eqs.~(\ref{Eq.1}) and (\ref{Eq.3}) predicts $g_A=1$.
The Lagrangian form in Eq.~(\ref{Eq.3})is more suitable to our discussion 
because it is similar to (\ref{Eq.1}) and shows explicitly an additional 
$S\pi NN$ vertex which is absent in (\ref{Eq.1}).
Equivalence between pseudoscalar and
pseudovector forms was verified explicitly in the case of TPEP \cite{Rob+95}
and in $\pi N$ form factor \cite{Mae+97}. More details of chiral scalar 
can be found in Refs.~\cite{Mae+98,Mae+97}.
It is worth noting that $S$ and $\sigma $ fields behave 
differently under chiral transformation. In the non-linear realization of chiral 
symmetry, the axial transformation of $S$ results in : $\delta ^AS=0$, while, 
in the linear realization, one has $\delta ^A\sigma =\vec \beta \cdot \vec \pi $, 
where $\vec \beta \,$ is an arbitrary isovector.

\begin{figure}
\epsfxsize=15.0cm
\centerline{\epsffile{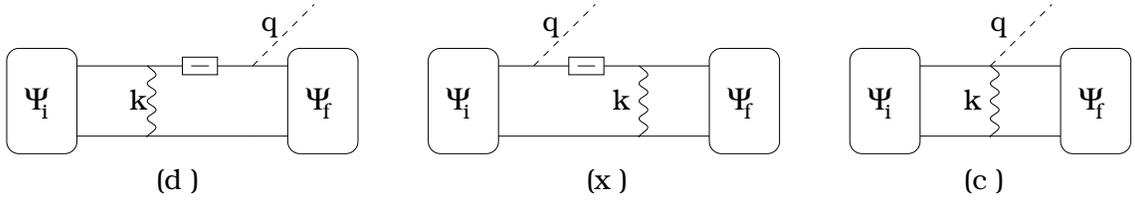}}
\vspace{1cm}
\caption{Diagrams with the chiral scalar $S$ (wavy line) that contribute
to the pion production (dashed line). The minus sign in the nucleon 
propagator (full line) indicates that the  positive energy states were 
removed. The labels means (d) for direct, (x) for crossed and (c) 
for contact.}
\label{Fig.1}
\end{figure}

Both Lagrangians in Eqs.~(\ref{Eq.1}) and (\ref{Eq.3}) 
have an isoscalar-nucleon and a pion-nucleon interaction
that give rise to the first two diagram of Fig.~\ref{Fig.1}d,x
for pion production process. However, the chiral $S$ Lagrangian has an extra
isoscalar-pion-nucleon interaction. This extra interaction vertex give raise
to the contact diagram depicted Fig.~\ref{Fig.1}c, which 
represents a genuine difference between $\sigma$-linear and chiral 
scalar $S$ lagrangians. 

As the first two diagrams of Fig.~\ref{Fig.1} 
are common to both Lagrangians, we set
a parameter $\lambda $ to turn on ($\lambda =1$) or off ($\lambda =0$) the
contributions of contact term and write 
\begin{equation}
T\left( \lambda \right) =
-igg_S^2\tau_c \left\{ \bar u\left( {\bf p}^{\prime}\right) \left[ 
\frac{-\not\! q}{p_d^2-m^2}+\frac{-\not\! q}{p_x^2-m^2}+
\frac{\lambda}{g\,f_\pi}\right] 
\gamma _5u\left( {\bf p}\right) \right\} ^{\left( 1\right) } \;
 \frac 1{k^2-m_s^2}\left[ \bar u\left( {\bf p}^{\prime }\right)
u\left( {\bf p}\right) \right] ^{(2)}+\left( 2\rightarrow 1\right) \,, 
\label{alambda}
\end{equation}
where $\tau_c $ is the isospin operator for the emitted pion, 
$\bar u\left( {\bf p}^{\prime}
\right) $ and $u\left( {\bf p}\right) $ are the spinors of nucleons with
momentum ${\bf p}^{\prime }$ and ${\bf p}$ respectively, $k=p_2^{\prime}
-p_2 $, $m$ and $m_S$ are the nucleon and scalar masses, 
the labels $(1)$ and $(2)$ denote nucleon $1$ and $2$, and the labels
$d$ and $x$ means direct and crossed diagrams.

Following the procedure applied in \cite{Mae+98}, we extract the irreducible
amplitude in order to avoid double counting due to positive frequency
propagation of nucleons; the result is the following proper amplitude for
the process $\pi N_jN_k$
\begin{eqnarray}
T_{\pi NN}\left( \lambda \right) &=&-igg_S^2\left\{ \tau_c\bar u\left( 
{\bf p}^{\prime }\right) \left[ \frac{\not\! q}{2E_d\left( p_d^0+E_d\right) }
+\frac{\not\! q}{2E_x\left( p_x^0+E_x\right) }
-\left( \frac 1{2E_d}-\frac 1{2E_x}\right) \gamma ^0+
\frac{\lambda}{gf_\pi}\right] \gamma _5u\left( {\bf p}\right) \right\} ^{(1)}  
\nonumber \\
&&\times \frac 1{k^2-m_s^2}\left[ \bar u\left( {\bf p}^{\prime }\right)
u\left( {\bf p}\right) \right] ^{(2)}+\left( 2\rightarrow 1\right) ,
\label{TvwpiNN}
\end{eqnarray}
where $E_{d\text{,}x}=\sqrt{{\bf p}_{d,x}^2+m^2}$ is the on mass-shell
nucleon energy.

\begin{figure}
\centerline{\epsffile{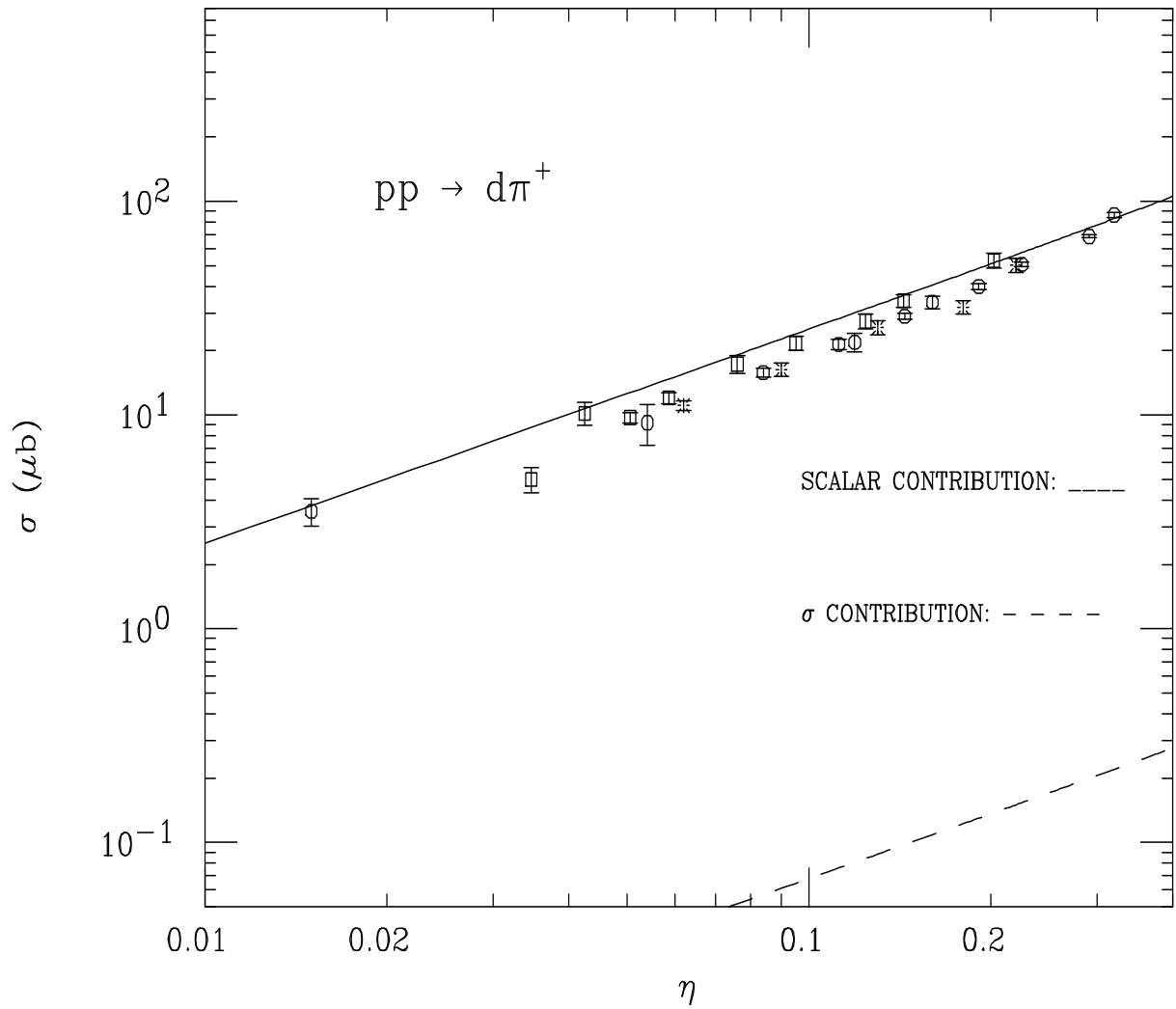}}
\vspace{-5.0cm}
\caption{Cross-section for the deuteron final state
as function of $\eta$. The graph show the contributions of chiral scalar
(full line) and the $\sigma$ meson (dashed line), where the potential 
used is Argonne v18. Data are from 
TRIUMF (squares) \protect\cite{Hut+91}, 
COSY (diamonds) \protect\cite{Dro+96}  and 
IUCF(circles) \protect\cite{Hei+96}. In this comparison, our results are obtained 
in the same fashion as done in Fig.~\ref{Fig.3}.}
\label{Fig.2}
\end{figure}

In order to obtain the production kernel we adopt the center of mass 
reference and take the non-relativistic limit
of the amplitude, Eq.~(\ref{TvwpiNN}), keeping terms up to
order $\omega /m$ where $\omega = E^{\prime } - E $. We obtain the following relations: 
\begin{eqnarray}
\frac 1{2E_d}-\frac 1{2E_x} &\sim &\frac{{\bf p}^2}{m^3}, \\
\frac 1{2E_d\left( p_d^0+E_d\right) }+\frac 1{2E_x\left( p_x^0+E_x\right) }
&\sim &\frac 1{2m^2},
\end{eqnarray}
\begin{eqnarray}
\bar u\left( {\bf p}^{\prime }\right) u\left( {\bf p}\right) &\rightarrow &%
\frac{\left( \bar E+m\right) }{2m}\left[ 1-\frac{{\bf \sigma }\cdot {\bf p}%
^{\prime }{\bf \sigma }\cdot {\bf p}}{\left( \bar E+m\right) ^2}\right] , \\
\bar u\left( {\bf p}^{\prime }\right) \gamma _5u\left( {\bf p}\right)
&\rightarrow &-\frac 1{2m}\left[ {\bf \sigma }\cdot \left( {\bf p}^{\prime }-%
{\bf p}\right) -\frac \omega {2\left( \bar E+m\right) }{\bf \sigma }\cdot
\left( {\bf p}^{\prime }+{\bf p}\right) \right] , \\
\bar u\left( {\bf p}^{\prime }\right) \gamma ^0\gamma _5u\left( {\bf p}%
\right) &\rightarrow &\frac 1{2m}\left[ {\bf \sigma }\cdot \left( {\bf p}%
^{\prime }+{\bf p}\right) -\frac \omega {2\left( \bar E+m\right) }{\bf %
\sigma }\cdot \left( {\bf p}^{\prime }-{\bf p}\right) \right] , \\
\bar u\left( {\bf p}^{\prime }\right) \not k\gamma _5u\left( {\bf p}\right)
&\rightarrow &\frac{\left( \bar E+m\right) }{2m}\chi ^{\dagger }\left[ \frac 
\omega {\left( \bar E+m\right) }{\bf \sigma }\cdot \left( {\bf p}+{\bf p}%
^{\prime }\right) -{\bf \sigma }\cdot {\bf k}\right] , \\
\frac 1{q^2-m_s^2} &\cong &-\frac 1{{\bf q}^2+m_s^2},
\end{eqnarray}
where  $\bar E=\left( E^{\prime }+E\right) $.
Using these results, and keeping only terms of the order ${\bf p}/m$, 
we get the non-relativistic amplitude:
\begin{equation}
t_{\pi NN}\left( \lambda \right) = -ig\frac{\omega_q}{2m}\frac 1{2m^2}\;
 \frac{g_S^2}{{\bf k}^2+m_S^2}
 \left\{
\tau _c^{\left( 1\right) }\left[ \left( 1+\frac \lambda 2\right) {\bf \sigma 
}^{\left( 1\right) }\cdot \left( {\bf p}+{\bf p}^{\prime }\right)
+\lambda \frac{2m}{\omega_q} {\bf \sigma }^{\left( 1\right) }\cdot
\left( {\bf p}^{\prime }-{\bf p}\right) \right] \right\}
+\left( 1\leftrightarrow 2\right)\,,
\label{tpp1}
\end{equation}
where $\mu$ is the pion mass and
$\omega_q=\sqrt{{\bf q}^2+\mu^2}$ is the on-shell emitted pion energy with
trimomentum ${\bf q}$ in the center of mass reference; 
${\bf k}={\bf p}-{\bf p}^{\prime}$ is the momentum transferred; 
${\bf \sigma }^{\left( i\right) }=\left\langle \chi ^i\left| 
{\bf \sigma }\right| \chi ^i\right\rangle $, $\tau _c^{\left( 1\right)
}=\left\langle \eta ^i\left| \tau _c\right| \eta ^i\right\rangle $,
$\,\left| \chi ^i\right\rangle $ and$\,\left| \eta ^i\right\rangle \,$ are
spin and isospin states of nucleon $i$ respectively. 
At this point it is possible to
figure out the main difference between chiral scalar $S$ and 
$\sigma $-linear. The pure $\sigma$ contribution gives (see 
Ref.~\cite{Coh+96}, Eq. (16)):
\begin{equation}
t_{\pi NN}^\sigma = -ig\frac{\omega_q}{2m}\frac 1{2m^2}\;
\frac{g_S^2}{{\bf k}^2+m_S^2} \left\{
\tau _c^{\left( 1\right) } {\bf \sigma 
}^{\left( 1\right) }\cdot \left( {\bf p}+{\bf p}^{\prime }\right)
 \right\}
+\left( 1\leftrightarrow 2\right)\;.
\label{t2}
\end{equation}
where we made use of GT relation. 
We can see that the contribution of $S$ \thinspace 
$\left( \lambda =1\right) $ adds  50\% in the 
term proportional to  $\left( {\bf p}+{\bf p}^{\prime }\right)$ and 
a new factor of 
$({2m}/{\omega_q})$ which multiplies the momentum exchanged. This new term 
will generate a derivative of the Yukawa function in the coordinate space, 
which is the same kind of contribution given by the Weinberg-Tomosawa term
\cite{KR66}.
We can see the consequence of this new term at threshold
$(\omega_q\approx\mu, \frac{2m}{\mu}\approx 13)$ in Fig.~\ref{Fig.2},
where we show the dominant role of chiral scalar for the reaction 
$pp\rightarrow d\pi^+$. 
We set $g_S=g_\sigma=8.7171$ and $m_S=m_\sigma=550$ MeV in order to compare
the results of both models. These parameters are 
obtained from table A.3 of Ref.~\cite{Mac89}, however, 
as $S$ is an effective particle both $g_S$ and $m_S$ are free
parameters. In principle one can extract the mass of the chiral scalar
by a careful analysis of the two  pion exchange role in the $NN$ potential.
This work is now in progress \cite{PR2000}.

The non-relativistic approach  to evaluate the 
cross section uses eigenstates $\left|NN\right\rangle $ 
of realistic $NN$ potentials to split the correlations
between nucleons from interactions which lead to pion production. Cohen 
{\it et al.} \cite{Coh+96} 
showed that the difference between the several realistic 
potentials available in the literature are minimal, except for the $\Delta$ contribution. Therefore we choose here 
to work with the Argonne v18 Potential \cite{WSS95}.
Thus, the
cross section reads
\begin{equation}
\sigma \propto \left\langle NN\right|\; |t_S|^2\;\left| NN\right\rangle \,,
\end{equation}
where $t_S$ is the pion production amplitude in the configuration space
due to the chiral scalar.
In order to calculate $t_S$, we follow Ref.\cite{HMG94} by including the 
effects of form factors defined by the following replacement:
\begin{equation}
g_S\to g_S{\Lambda_S^2-m_S^2\over\Lambda_S^2-{\bf k}^2},
\label{FF}
\end{equation}
where ${\bf k}$ is the transferred momentum and $\Lambda_S=2000$ MeV is the 
cutoff mass, listed in Table A.3 of Ref.~\cite{Mac89}. 
The final amplitude in the coordinate space at threshold reads:

\begin{equation}
t_{S}\left( \lambda \right) =-i\frac g{2m}\frac{g_S^2}{4\pi }\frac 
\omega {2m}\left\{ \frac 1mF_1\left( r,\Lambda_S \right) {\bf \Sigma}_c\cdot
2{\bf p}-\frac 1mi{\bf \Sigma }_c\cdot {\bf \hat r}F_2\left( r,\Lambda_S
\right) 
+\frac \lambda {2m}F_1\left( r,\Lambda \right) {\bf \Sigma}
_c\cdot 2{\bf p}-\frac{\lambda}{2m}\left( 1-\frac{4m}\omega \right) i{\bf 
\Sigma}_c\cdot {\bf \hat r}F_2\left( r,\Lambda_S \right) \right\}\,,
\label{tS}
\end{equation}
where 
\begin{eqnarray}
F_1\left( r,\Lambda \right) &=&\left( \frac{e^{-m_sr}}r-\frac{e^{-\Lambda r}}
r-\frac 1{2\Lambda }\left( \Lambda ^2-m_S^2\right) e^{-\Lambda r}\right) , \\
F_2\left( r,\Lambda \right) &=&\left( -\left( 1+m_sr\right) \frac{e^{-m_sr}}{
r^2}+\left( 1+\Lambda r\right) \frac{e^{-\Lambda r}}{r^2}+\frac 12\left(
\Lambda ^2-m_S^2\right) e^{-\Lambda r}\right)\,,
\end{eqnarray}
and ${\bf \Sigma}_c \equiv\left( \tau _c^{(1)}{\bf \sigma }
^{(1)} -\tau _c^{(2)}{\bf \sigma }^{(2)} \right)$. 

The next step is the evaluation of the matrix element of the operator
in Eq.~(\ref{tS}) between the initial and final states for the 
reactions  $pp\rightarrow d\pi^+$, $pp\rightarrow pn\pi^+$ and
$pp\rightarrow pp\pi^0$. We limit ourselves to the threshold region, where 
the two final state nucleons are in a strong attractive $L_{NN}=0$ state, and
we can consider the pion angular momentum $\ell_\pi=0$. The last step is to square the
amplitudes and integrate over the phase space avaliable.
These calculations are straightforward and can be found in \cite{Coh+96}
for the $pp\rightarrow pp\pi^0$ reaction and in \cite{RMK99} for the other
two reactions.

\begin{figure}
\centerline{\epsffile{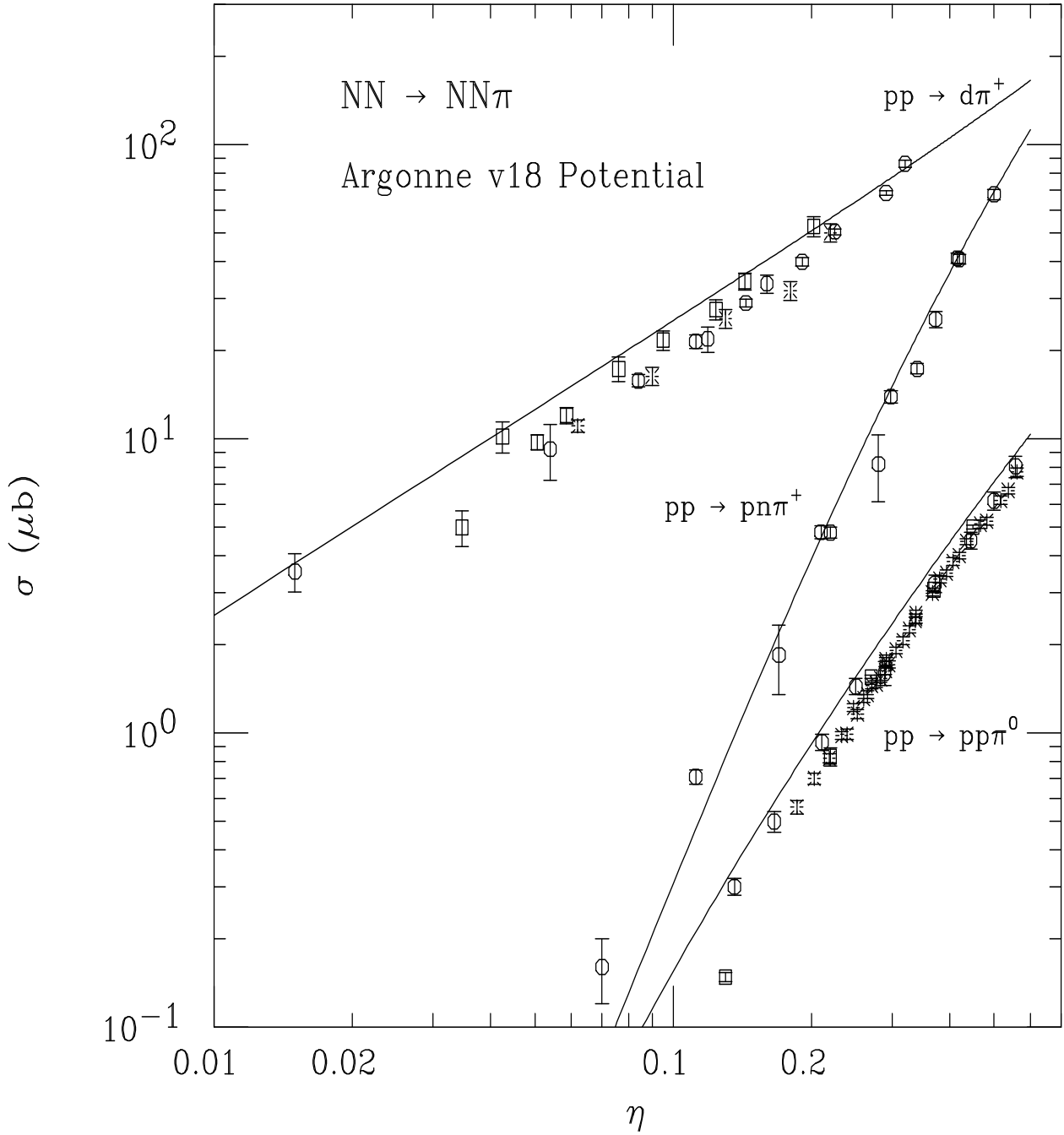}}
\vspace{-5.0cm}
\caption{Cross-section $\sigma$ for the pion production 
as function of $\eta$. The lines shown the contribution 
of the chiral $S$ scalar for the reactions 
$pp\rightarrow pp\pi^0$, $pp\rightarrow d\pi^+$, and 
$pp\rightarrow pn\pi^+$ using the Argonne V18 potential.
The $pp\rightarrow pp\pi^0$ data are from 
\protect\cite{Mey+90} (circles) and \protect\cite{Bon+95}
(crosses); the $pp\rightarrow d\pi^+$ data are from
\protect\cite{Hut+91} (crosses), \protect\cite{Dro+96}
(squares) and \protect\cite{Hei+96} (circles);
the $pp\rightarrow pn\pi^+$ data are from 
\protect\cite{Dae+95,Dae+98} (circles).}
\label{Fig.3}
\end{figure}

We show our results in Fig.~\ref{Fig.3}, 
where we plot the cross sections for the 
reactions $pp\rightarrow pp\pi^0$, $pp\rightarrow d\pi^+$, 
and $pp\rightarrow pn\pi^+$
as a function of the pion momenta $\eta=\frac{q}{\mu}$. The impressive 
agreement between the chiral $S$ scalar contribution and the data 
reveals that this is
the major contribution for the pion production at threshold.
Since this chiral $S$ simulates two pion exchanges, this
result corroborates that the pion production reaction is a suitable framework
to study the medium and short ranged dynamics of NN interaction and, in some sense,
contributions due to long ranged terms have a minor role in these reactions. 
It is worth noting that this work do not intend to fit the data, but only show 
that we do not have the old factor of 5 below data anymore. A
detailed study comparing this chiral $S$ and long ranged terms will be shown
elsewhere \cite{MR2000}, but we present in Tab.~\ref{Tab.01} some preliminary results for 
the $pp\rightarrow d\pi^+$ cross section. We see in the table that the $S$ 
contribution is predominant and is due to the $S\pi NN$ vertex term, as seen in the
Fig.~\ref{Fig.1}c. When we turn this contribution off ($\lambda=0$ in Eq.~(\ref{tS})), 
the results of $S$ and $\sigma$ are identical. This means that the direct and crossed diagrams
presented in both contributions are of order $\mu/m$ lower in comparison with the 
$S\pi NN$ term. In addition, the table shows that the contributions of IA, $\Delta$ and SEA 
are of order $\mu/m$ lower than the $S$ term, and the WT term is an exception, due to the 
isospin coefficients. 

\begin{table}
\caption{Prelimirary $\sigma/\eta$ results for the reaction $pp\rightarrow d\pi^+$
for $\eta=0.01$, where the abbreviations read as: WT: Weinberg-Tomosawa, IA: impulse approximation, $\Delta$: delta resonance contribution, GC: Galilean correction to WT, 
$\sigma$: $\sigma$ linear meson contribution, and SEA: $\pi\pi NN$ isoscalar contribution. 
All these terms except the $S$ contribution can be seen in Refs.~\protect\cite{Coh+96,RMK99}.}
\begin{tabular}{lcccccccr}
Contribution &
& Scalar & WT  & GC   & IA   & $\Delta$ & $\sigma$ & SEA \\ \tableline
$\sigma/\eta \;\;(\mu b)\;$ & 
& 262    & 110 & 3.72 & 0.39 &    1.24  &    0.18  & 0.64 \\
\end{tabular}
\label{Tab.01}
\end{table}

Our results for the $S$ contribution to the pion production 
have shown that the cross section for these reactions
near threshold can be explained by taking into account the contribution of the 
medium-range three-pion contact interaction with the nucleon, 
here represented by the $S\pi NN$ vertex. 
In a $\chi_{PT}$ approach \cite{Coh+96}, this represents that the second order contributions,
expressed mainly by the two-pion loops, are the most important contributions.
It could be an indication of a connection between the energy
threshold of  a particular reaction and the dominant order of the power
counting. Since the chiral $S$ scalar simulates the
two-pion exchange in the pion production, one important issue here is to
check if the sum of all contributions of second order in  $\chi_{PT}$ gives
the same result obtained here. This is a herculean task, but some promising
results were shown recently \cite{Dmi+99}.

To conclude, we have shown that the cross section for the reactions 
$pp\rightarrow pp\pi^0$, $pp\rightarrow d\pi^+$, and $pp\rightarrow pn\pi^+$ 
near threshold can be explained by taking into account the contribution of the 
medium-range three-pion contact interaction with the nucleon, 
here represented by the $S\pi NN$ vertex, to the pion production amplitude; 
in a $\chi_{PT}$ approach, this fact means that the second order contribution,
expressed mainly by the two-pion loops, is the most important contribution.
We believe that should be a close connection between the energy threshold of
a particular reaction and the dominant order of the power counting. This idea 
may be not new, but the present results can be understood using this 
different approach.

\acknowledgements
We thank M.R. Robilotta and U. van Kolck for helpful discussions. This work was supported by 
FAPESP Brazilian Agency under contract numbers 97/6209-4 (C.A.dR.) and 99/00080-5 (C.M.M.).
One of us (C.A.dR.) thanks the Nuclear Theory Group at University of Washington for 
the hospitality during the initial stages of this work.

\end{document}